\documentclass[12pt]{iopart}

\usepackage{iopams,amssymb,graphicx,color,bm,epstopdf,dcolumn}

\begin{document}

\title[]{Giant Rabi splitting in metallic cluster - cavity system}

\author{I G Savenko,$^{1}$ R G Polozkov,$^{2}$ and I A Shelykh$^{1,3}$}

\address{$^{1}$Science Institute, University of Iceland, Dunhagi 3, IS-107, Reykjavik, Iceland}
\address{$^{2}$St.Petersburg State Polytechnical University, Politechnicheskaya 29, 195251 St.Petersburg, Russia}
\address{$^{3}$International Institute of Physics, Av. Odilon Gomes de Lima, 1722, Capim Macio, CEP: 59078-400, Natal- RN, Brazil}

\ead{Ivan.G.Savenko@gmail.com}

\begin{abstract}
We investigate theoretically the photoabsorbtion cross-section of a cluster of
alkali atoms embedded in a single-mode quantum microcavity. We
show that if the energy of the giant plasmonic resonance lies
close to the energy of the cavity mode, the strong coupling between
plasmon and cavity photon can occur which is characterized by mode
anticrossing and observation of the doublet structure in
ptotoabsorbtion spectrum. The characteristic values of the Rabi
splitting are expected to be several orders of magnitude larger than
those observed in single quantum dot - cavity systems.
\end{abstract}


\maketitle


\section{Introduction}

Strong coupling between light and matter
excitations attracts growing interest of the physical community.
The problem is important not only because of the fundamental aspects
brought forward by the interaction of material systems with photons
\cite{mabuchi02a}, but also due to the possibility of use the
strong coupling phenomena for creation of optoelectronic devices
of the new generation, such as polariton lasers \cite{Christopoulos},
optical logic gates \cite{Leyder}, all-optical integrated circuits
\cite{Liew2008}, terahertz light sources \cite{Savenko2011} and others.

Coupling of a zero-dimensional (0D) quantum system to a single photon
mode which forms the subject of cavity Quantum Electrodynamics
(cQED), is of particular importance from this point of view due to potential application of cQED to quantum information
processing \cite{imamoglu99a,bennett00a,Johne2008}. In the domain of
the condensed matter, the system which recently attracted particular
attention consists of a quantum dot (QD) coupled to a single
microcavity mode \cite{andreani99a,reithmaier04a,yoshie04a,peter05a,kaliteevski07a}.
The material excitations in the QD are excitons - bound
electron-hole pairs. Owing to their spatial confinement and energy
level discretization, they can be brought in strong coupling with
the mode of a microcavity: if a resonant absorber of light is embedded in a microcavity it results in a new regime of mixed light modes. The condition of the strong coupling of the absorber and the cavity is that the light-matter interaction- induced splitting, known as the Rabi splitting, is greater than the spectral linewidths (inverse lifetimes) of both the photon and exciton modes.
There exists many examples besides QDs, such as: a pillar (etched planar cavity)\cite{reithmaier04a}, the defect of a photonic
crystal \cite{yoshie04a} or the whispering gallery mode of a
microdisk \cite{peter05a,kaliteevski07a}, among others.
In Refs.~\cite{reithmaier04a,yoshie04a,peter05a} such
structures have demonstrated Rabi doublet in their optical
spectra which characterises the mode anticrossing that marks
the overcome of dissipation by the coherent exciton-photon
interaction. The characteristic values of Rabi splitting for QDs are typically  of the order of 10-100 $\mu$eV which is 3-4 orders of magnitude smaller than for the planar microcavities with quantum wells as an active region. Small
value of the Rabi splitting makes achievement of the regime of strong coupling in 0D systems technically complicated task
and limits possibilities of its practical implementations.

A natural question arises: can strong coupling regime be observed
for material excitations other than excitons? One of the main
candidates is collective plasmonic excitation in metals. For planar metallic
structures the dispersion of the surface plasmon lies outside the
light cone that makes its direct optical excitation impossible
\cite{Pitarke2007} and rules out the possibility of the observation
of any strong coupling effects. This is not true, however, for more
complicated structures containing metallic nanowire arrays
\cite{Christ2003} and metallic nanorods \cite{Huang2010} for which
the effects of strong coupling were shown to be extremely pronounced
and experimentally observed Rabi splitting can be as big as 250 $\mu$eV
\cite{Christ2003}. This number exceeds by the order of magnitude
the values of Rabi splitting for exciton-photon coupling in planar
inorganic microcavities \cite{KavokinBook} and is comparable to Rabi
splittings observed in organic structures \cite{Hobson2002}.

In the present manuscript we consider another type of hybrid metal-dielectric
structure consisting of the single metallic cluster
embedded inside a single-mode photonic cavity. The collective motion
of electrons in the cluster against the positively charged ionic
background leads to formation of a surface plasmonic mode
responsible for the appearance of a giant resonance in the
photoabsorbtion spectra \cite{deHeer1993,Brack1993,Brechignac1994,Madjet1995,Ekardt,Solovyov}. If the energy of the plasmon is close to that of the photonic mode of the cavity,
the processes of multiple resonant emissions and absorptions of
photons by the cluster take place and hybrid plasmon-photon
modes are being formed. The goal of the present manuscript is to analyze their
influence on photoabsorbtion spectrum of the system and to estimate the corresponding
values of the Rabi splitting.


\section{Model}

\begin{figure}[tbp]
\includegraphics[width=1.0\linewidth]{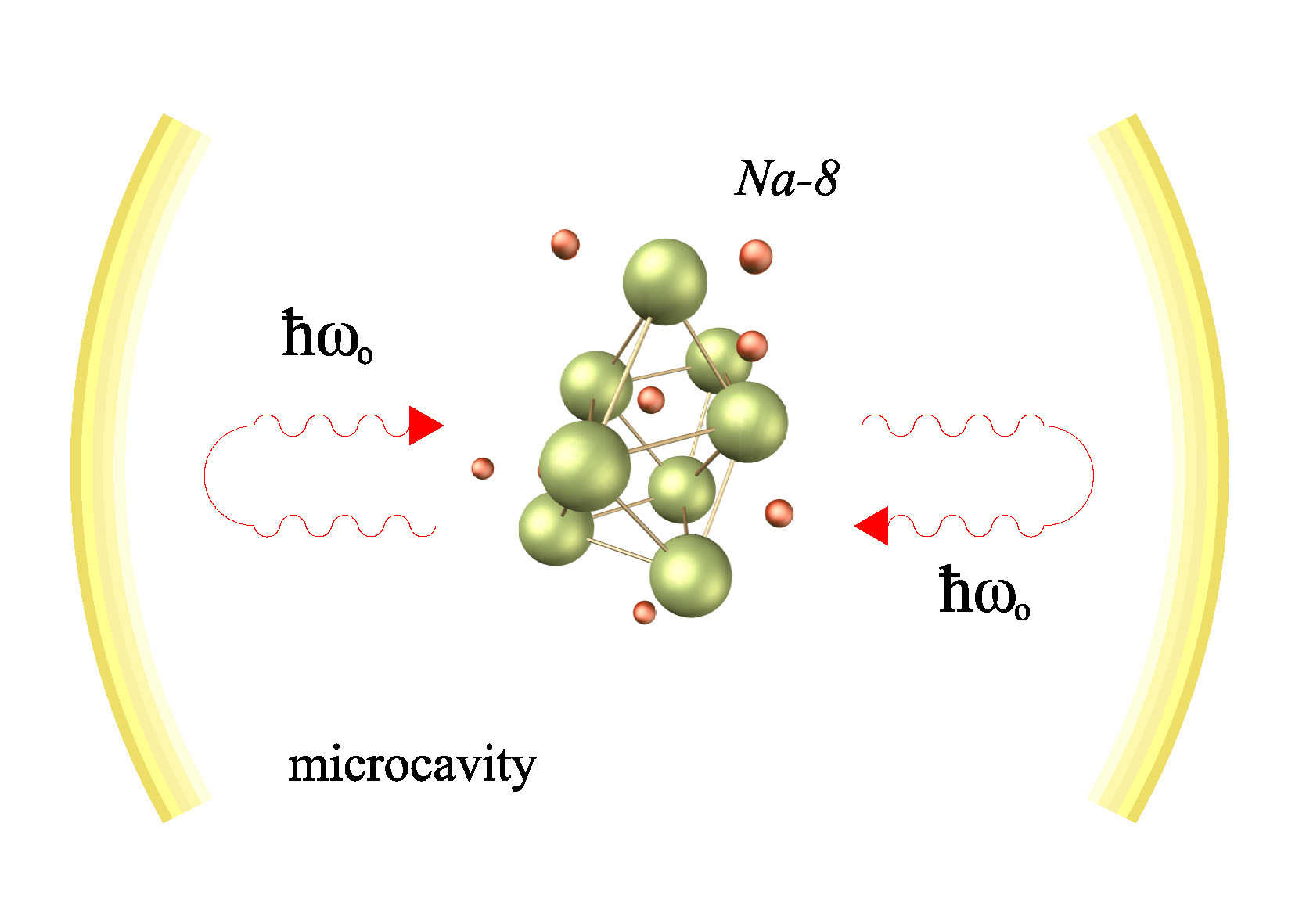}
\caption{System geometry: metallic cluster formed by alkali atoms Na-8 embedded in a single-mode microcavity. The cavity photons of the frequency $\omega_0$ undergo multiple re-emissions and re-absorbtions by the collective excitation of the cluster forming light-matter hybrid polariton eigenmodes of the system.}
\label{AlkaliCluster}
\end{figure}

To illustrate the onset of the strong coupling regime in cluster - cavity system we will focus on the
metallic clusters formed by alkali atoms (Li, Na, K), as they
are the most exhaustively studied from both theoretical and experimental
points of view. The geometry of the system we consider is shown in Fig.~\ref{AlkaliCluster}

The theoretical basis of studying the collective
phenomena in clusters is provided by the jellium model in which $N$
electrons (one per each alkali atom) are moving in the spherically
symmetric electrostatic potential $V_b(r)$ formed by the uniform
background of the positive ions calculated as convolution of the
Coulomb electrostatic potential with smooth charge distribution
\begin{eqnarray}
V_b(r)=\left\{\begin{array}{cc}-\frac{Ne^2}{8\pi\varepsilon_0R}\left[3-\left(\frac{r}{R}\right)^2\right],r\leq
R,\\
-\frac{Ne^2}{4\pi\varepsilon_0r}, r>R.\end{array}\right.
\end{eqnarray}
 This assumption is generally believed to be well justified for the
 alkaline clusters with closed shells \cite{Brack1993}, i.e. for $N=8,18,20,34,40,58,92,...$
 by reason of applicability of two main conditions: first, the valence electrons must be strongly delocalized. And second, the valence electrons should have an s-wave character with respect to the ionic core. However, we want to underline that our results are of general character and should be qualitatively the same for clusters with partially filled shells and for clusters consisting of the non-alkali atoms, but he computation procedure becomes more tricky in these latter cases \cite{Thermes}.

Let us consider an alkali-metal cluster with transition frequencies $\omega_{\alpha j}=\omega_j-\omega_\alpha=\hbar^{-1}(E_j-E_\alpha)$ where the indices $\alpha$ and $j$ correspond to the occupied and unoccupied single electron states respectively. In the case when electron-electron interactions are neglected or treated within the Hartree-Fock approximation, each single electron transition is coupled to the external photons independently. As the oscillator strengthes of these individual transitions are small, their coupling with the cavity mode results in small values of the Rabi splitting (of the order of magnitude of those observed in QD-cavity systems).

However, the account of many-body corrections can change the picture dramatically. Coulomb interaction leads to the formation of the collective excitation of the cluster known as the Giant plazmonic resonance and has a huge oscillator strength (several orders of magnitude larger than oscillator strengthes of individual transitions) \cite{deHeer1993,Brack1993,Brechignac1994,Madjet1995,Solovyov}. Consequently, one can expect high values of the Rabi splitting in the coupled cluster - cavity system.


\section{Formalism}
The interaction of the electromagnetic field with material objects is described by dipole matrix elements of the transitions $d_{\alpha j}$. The account of many electron processes results in renormalization of the dipole matrix elements which become frequency-dependent. They are denoted as $D^{(0)}_{\alpha j}(\omega)$ in our further discussion. Account of processes with photons in microcavity leads to further renormalization, $D_{\alpha j}(\omega)$.

\begin{figure}[tbp]
\includegraphics[width=1.0\linewidth]{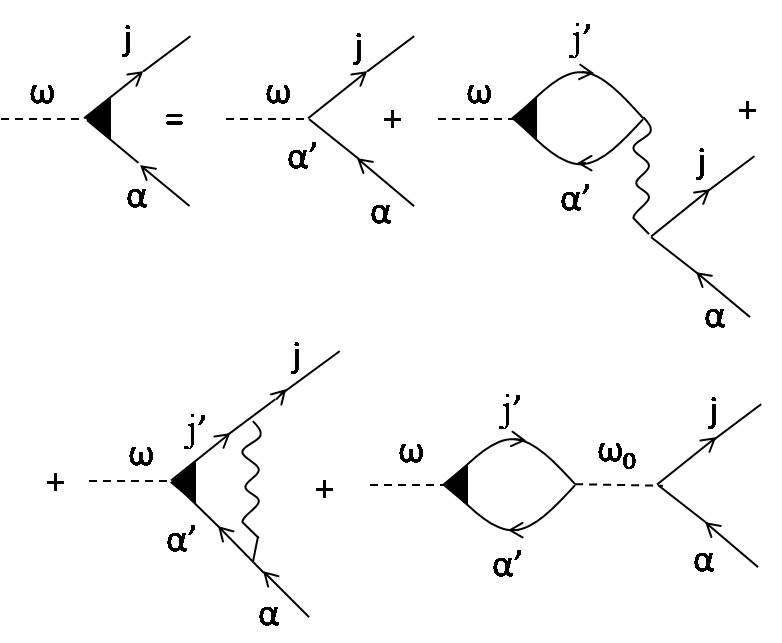}
\caption{Diagrammatic representation of the equation for the renormalized dipole matrix element of the transition. The straight lines with arrows correspond to the electrons in the cluster at vacant and filled orbitals $j,\alpha$, the wave lines to the Coulomb interaction, the doted lines - to the cavity photons of frequency $\omega_0$.}
\label{FeynmannDiagramme1}
\end{figure}

\begin{figure}[tbp]
\includegraphics[width=1.0\linewidth]{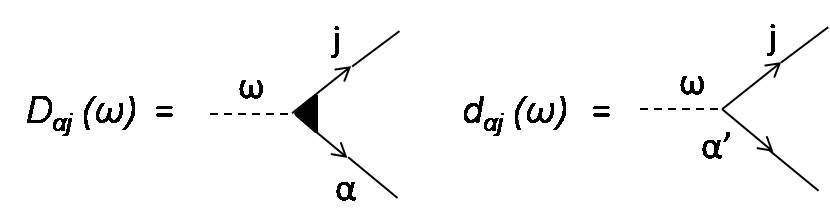}
\caption{Diagrammatic representation of renormalized and bare dipole matrix elements of the transition.}
\label{FeynmannDiagramme2}
\end{figure}

Figure~\ref{FeynmannDiagramme1} illustrates the diagrammatic representation of the equation for $D_{\alpha j}(\omega)$ in the system we consider, and graphical representations of renormalized and bare dipole matrix elements of photoabsorbtion $D_{\alpha j}(\omega)$ and $d_{\alpha j}$ are presented in Fig.~\ref{FeynmannDiagramme2}. On the right hand side of Fig.~\ref{FeynmannDiagramme1} the first diagram corresponds to the direct photoabsorbtion, the second two diagrams account for
the influence of dynamical polarizability in the random phase approximation with exchange (RPAE) and give rise to giant plasmon resonance, and the last one accounts for the multiple re-emissions and re-absorptions of the cavity photon. In figures the straight lines with arrows correspond to the electrons in the cluster at vacant and filled orbitals $j,\alpha$, the wave lines --- to the Coulomb interaction, the dotted lines --- to the cavity photons of frequency $\omega_0$.
Using Goldstone technic of evaluation of the Feynmann diagrams one can find the renormalized dipole matrix elements from the following set of  non-linear algebraic equations
\begin{equation}
D_{\alpha j}(\omega)=d_{\alpha j}(\omega)+
\left(\sum_{\alpha'\leq F \atop j'> F}-\sum_{\alpha' > F \atop j'\leq F}\right)
\frac{M_{\alpha,\alpha';j,j'}{(d,D(\omega))}D_{\alpha' j'}(\omega)}{\omega \mp \omega_{\alpha'j'}\pm i\delta},
\label{Dyson}
\end{equation}
where the first and second sum correspond to the 'time forward' and 'time backward' diagramms respectively and
\begin{eqnarray}
\label{Matrix}
\nonumber
M_{\alpha,\alpha';j,j'}(d,D(\omega))&=&\langle \alpha',j|V_C|j',\alpha\rangle-\langle j,\alpha'|V_C|j',\alpha\rangle-\\
&-&\frac{2\omega_0g_{\alpha' j'}^{\ast}(d)g_{\alpha j}(d)}
{\left(\omega^2-\omega_0^2+2i\omega_0\gamma\right)} \label{M}.
\end{eqnarray}
In the formula above $\langle |V_C|\rangle$ are the matrix elements of the exact Coulomb interaction with both the direct and exchange parts. The linewidths $\delta$ and $\gamma$ correspond to the finite lifetime of the one-electron excitations and cavity photons, $g_{\alpha j}$ are the matrix elements of the interaction between the transitions $\alpha\rightarrow j$ and cavity mode, which are proportional to the dipole matrix elements of the transitions and can be estimated as \cite{ScullyBook}
\begin{eqnarray}
g_{\alpha j}(d)=\omega_{\alpha j}\sqrt{\frac{\hbar}{2\varepsilon_0\omega_0V}}d_{\alpha j}\approx\omega_{\alpha j}\omega_0\sqrt{\frac{\hbar}{2\pi^3\varepsilon_0c^3}}d_{\alpha j},
\label{g}
\end{eqnarray}
%
where $V\approx(\lambda_0/2)^3$ denotes the cavity volume.

Note, that solving the system of equations (\ref{Dyson}) is equivalent to accounting the Coulomb RPAE diagrams and coupling between transitions in cluster and cavity mode up to the infinite order. Together, they describe the formation of the hybrid plasmon - photon excitations in the cluster - cavity system.

\begin{figure}[tbp]
\includegraphics[width=1.0\linewidth]{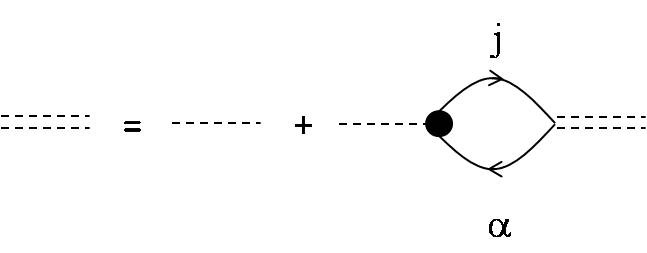}
\caption{Diagrammatic representation of Dyson equation for the Green's function of the cavity photon.The double dashed line corresponds to the renormalized Green's function of the photon, the single dot line --- to the Green's photon of the bare photon. The dot on the diagram corresponds to the renormalized matrix element of the cluster - cavity coupling accounting for many - body interactions in the cluster but neglecting multiple re-emissions and re-absorbtions.}
\label{FeynmannDiagramme3}
\end{figure}

The response of the system which can be described in terms of the photoabsorption cross-section is proportional to the imaginary part of the dipole dynamical polarizability
\begin{equation}
\sigma(\omega)=4\pi\frac{\omega}{c}Im\left[\alpha(\omega)\right]
\end{equation}
%
which can be found using standart formula
\begin{eqnarray}
\alpha(\omega)=-e^{2}\sum_{\alpha j}\left[\frac{\vert D_{\alpha j}(\omega)\vert ^{2}}{ \omega - \omega_{\alpha j}+i\delta}-\frac{\vert D_{\alpha j}(\omega)\vert ^{2}}{ \omega + \omega_{\alpha j}-i\delta}\right].
\end{eqnarray}
The energies of the eigenstates of the system can be found from the poles of the dressed Green's function for the cavity photon represented diagramatically in Fig.~\ref{FeynmannDiagramme3}. The double dashed line corresponds to the renormalized Green's function of the photon, the single dot line --- to the Green's function of the bare photon $G^0=2\omega_0/(\omega^2-\omega_0^2+2i\omega_0\gamma)$. The symbol \emph{dot} in the diagram corresponds to the renormalized matrix element of the cluster - cavity coupling $g_{\alpha j}[D^{(0)}(\omega)]$, accounting for many - body interactions in the cluster but neglecting multiple re-emissions and re-absorbtions of the cavity photon as latter processes are already accounted for in the diagram of the Dyson equation for the dressed photon (Fig.~\ref{FeynmannDiagramme3}) and should not be counted twice. Mathematically, the values of $D^{(0)}_{\alpha j}(\omega)$ can be found from the system of the equations analogical to those represented in diagrammatic form in Fig.~\ref{FeynmannDiagramme1} but with the last diagram on the right hand side being absent. The Green's function of the dressed photon reads
\begin{eqnarray}
G_{ph}=\frac{1}{(G_{ph}^0)^{-1}-\Sigma(\omega)},
\label{Gph}
\label{DysonPhoton}
\end{eqnarray}
where
\begin{equation}
\Sigma(\omega)=\sum_{\alpha j}\frac{g^*_{\alpha j}(d)g_{\alpha j}[D^{(0)}(\omega)]}{\omega-\omega_{\alpha j}+i\delta}.
\end{equation}
The eigenfrequancies of the hybrid eigenmodes can be found from the poles of expression (\ref{Gph}) giving the following transcendent equation:
\begin{eqnarray}
\omega^2-\omega_0^2+2i\omega_0\gamma-2\omega_0\Sigma\left(\omega\right)=0
\label{Eigen}
\end{eqnarray}

Before we proceed with presenting the results of the numerical modelling, let us
analyze the limiting cases.

On the one hand, if the cavity is absent, electron-electron interactions renormalize dipole matrix elements of the transitions $D^{(0)}_{\alpha j}(\omega)$ which become frequency dependent and form sharp peaks at the frequency $\omega=\omega_{p}$  corresponding to the characteristic frequency of the giant plasmon resonance \cite{deHeer1993,Brack1993,Brechignac1994,Madjet1995,Solovyov}.
On the other hand, if one neglects the Coulomb interactions (which corresponds to the retaining of the first and last diagrams only in the diagrammatic equation presented in Fig.~\ref{FeynmannDiagramme1}) and considers the coupling of the individual single transition to the cavity mode, the equations for the renormalized dipole matrix element and dressed photonic Green's function
\begin{equation}
D_{\alpha
j}(\omega)=\frac{d_{\alpha
j}\left(\omega^2-\omega_0^2+2i\omega_0\gamma\right)\left(\omega-\omega_{\alpha
j}+i\delta\right)}{\left(\omega^2-\omega_0^2+2i\omega_0\gamma\right)\left(\omega-\omega_{\alpha
j}+i\delta\right)-2g^2\omega_0}, \label{DSimple}
\end{equation}
\begin{equation}
G_{ph}=\frac{2\omega_0\left(\omega-\omega_{\alpha
j}+i\delta\right)}{\left(\omega^2-\omega_0^2+2i\omega_0\gamma\right)\left(\omega-\omega_{\alpha
j}+i\delta\right)-2g^2\omega_0}
\label{SimpleDyson}
\end{equation}
can be easily solved. The poles of these expressions determine the energies of the new hybrid eigenstates of the system. The condition of achievement of the strong coupling regime characterized by the mode anticrossing at resonance ($\omega_0=\omega_{\alpha j}$) is given by a standard expression $4g^2>(\delta-\gamma)^2$ \cite{Laussy2008}.


\section{Results and discussion}

\begin{figure}[tbp]
\includegraphics[width=1.0\linewidth]{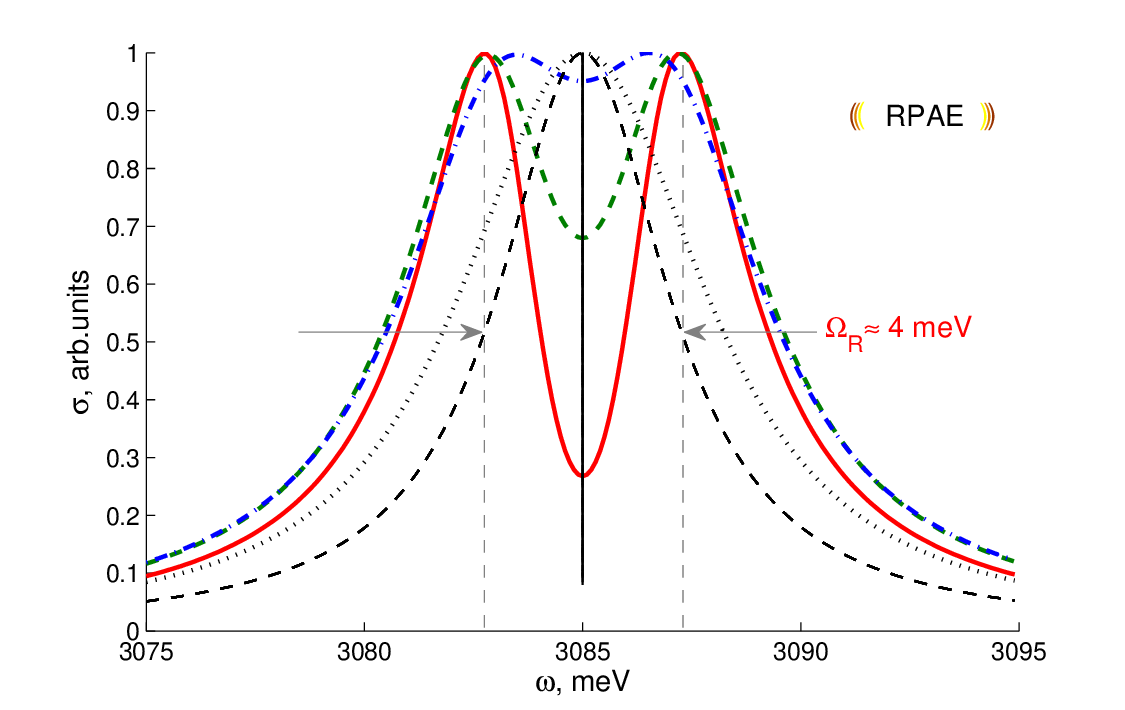}
\caption{Cluster Na-21$^+$ photoabsorption cross-section spectrum - system response on the exciting emittion of light for different lifetimes of the photons (in meV): 1 (red/solid) - quasi-ideal microcavity, 2 (green/dashed), 4 (blue/dash-dotted) and 6 (black/dotted) - limiting case which is close to absense of the cavity. Rabi splitting $\Omega_R$ decreases at the photons lifetime being decreased and finally desappears at $\gamma\approx 7$ meV.
See text for details.}
\label{FigResponse}
\end{figure}

Let us present now the results of numerical modeling of the realistic cluster - cavity systems. 
We consider Na -8, 9$^+$, 21$^+$, 42$^{++}$ clusters embedded in a photonic cavity in the position where the electric field of the cavity mode reaches its maximum (we explain why we chose ionic clusters below). The light mode structure in a microcavity is a standing wave ($sin$-like oscillating) with nodes at some coordinates. The typical size of a cavity (and light wavelength there also) is micrometers, while the size of a cluster is Angstroms. Thus we can approximately assume that the cluster is embedded in some local point in the cavity. If one puts a cluster in a node of the wave, the interaction of the matter mode with the light mode is absent but can be increased if move the cluster towards the anti-node. The stronger the coupling, the more pronounced the effect and Rabi splitting. The energy of the cavity mode $\omega_0$ is supposed to lay close to the energy of the giant plasmon resonance $\omega_{pl}$. The energies of single electron states in the cluster were calculated within the jellium model using Hartree - Fock approximation. The non-radiative linewidths of all the individual transitions in the cluster were taken to be the same and corresponding to $\delta=1$ meV for the cross-sections in Fig.~\ref{FigResponse}. 

\begin{figure}[tbp]
\includegraphics[width=1.0\linewidth]{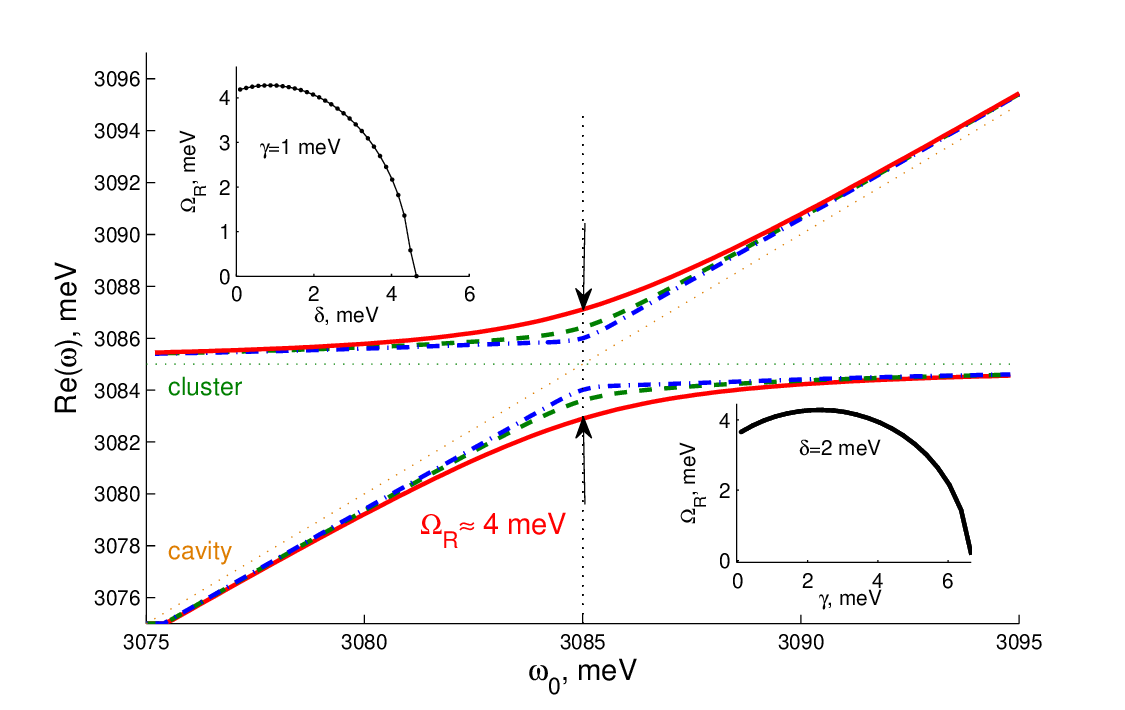}
\caption{Eigen frequency real part dependence on the microcavity mode wavelength for the Na-21$^+$ cluster for different lifetimes of the photons (in meV): 1.0 (red/solid), 4.0 (green/dashed) and 6.0 (blue/dash-dotted). Illustration of the crossover from strong- to weak-coupling regime. When $\gamma \approx 7$ meV the splitting $\Omega_R\approx 0$ and one can see two solutions corresponding to the bare photon and plasmon $\omega_1=\omega_0$ and $\omega_2=\omega_{cl}$ (dotted curves named 'cavity' and 'cluster'). Insets show dependences of the Rabi splitting $\omega_R$ on the inverse lifetimes of the photon $\gamma$ and plasmon $\delta$. The maxima on the dependences correspond to optimal combinations of the parameters (see discussion in the end of Section 'Formalism').}
\label{FigSplitting}
\end{figure}
\begin{figure}[tbp]
\includegraphics[width=1.0\linewidth]{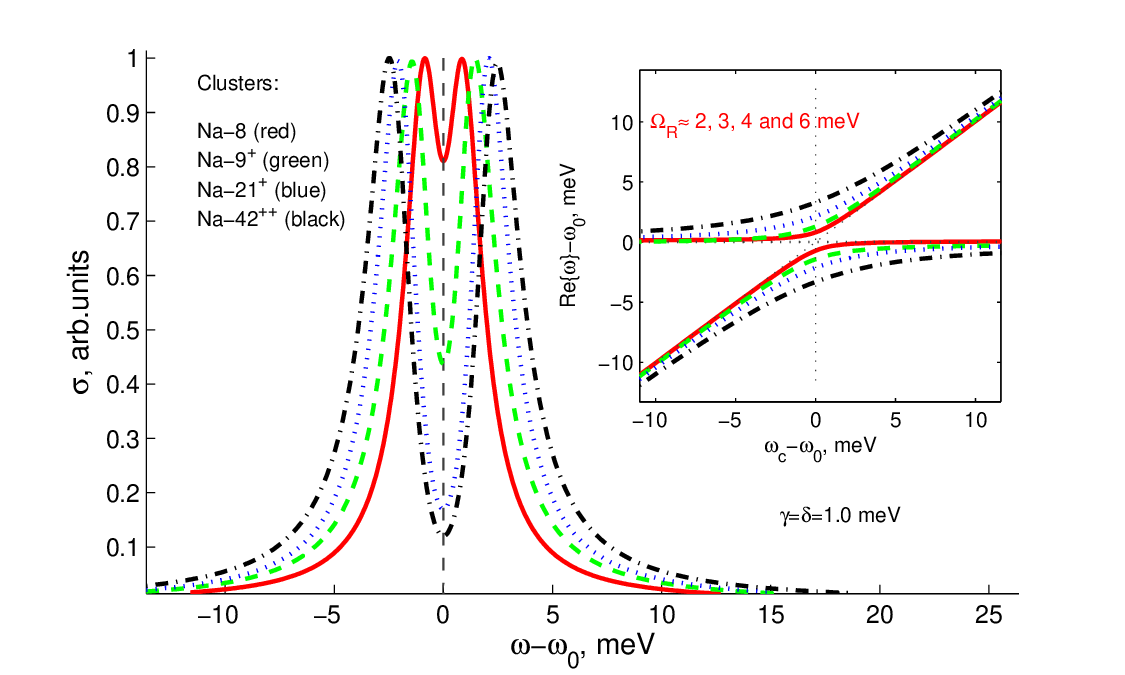}
\caption{Photoabsorption cross-section spectrum for different sodium clusters: Na-8 (red/solid) Na-9$^+$ (green/dashed), Na-21$^+$ (blue/dotted) and Na-42$^{++}$ (black/dash-dotted) as a function of 'shifted' with respect to the central frequency energy, $\omega-\omega_0$. Rabi splitting $\Omega_R$ varies from $\approx$ 2 to 6 meV with the increase of number of atoms in the cluster. Inset illustrates anticrossing for different clusters (the numerical solutions of the Dyson equation for the photon Green's function, see Eq.~\ref{DysonPhoton}-\ref{Eigen}). The inverse lifetimes of the photon and plasmon modes,$\gamma$ and $\delta$, were taken equal to 1 meV.}
\label{FigClusters}
\end{figure}

Figure~\ref{FigResponse} illustrates photoabsorbtion cross-section of the cavity - cluster system as a function of frequency of the external laser excitation for the case when the energy of the giant plasmon is tuned in resonance with the energy of the photonic mode, $\omega_0=\omega_{pl}$. If the lifetime of the photons in microcavity $\tau$ is low ($\gamma\approx\hbar/\tau$ is high) we can observe one peak on the photoabsorption spectrum (see Fig.~\ref{FigResponse}, note, all the plots are normalized to unity). However, the situation qualitatively changes if the lifetime increases. In this case the strong coupling regime is being established and two peaks appear in photoabsorbtion spectrum. The distance between the peaks corresponds to the Rabi splitting and increases with the increase of the lifetime of the cavity mode, it can reach the values of about 1.8, 2.1, 4.2 and 6.9 meV for Na-8, Na-9$^+$, Na-21$^+$ and Na-42$^{++}$ clusters respectively. These values are several orders of magnitude larger than those observed in single QD - cavity systems \cite{andreani99a,reithmaier04a,yoshie04a,peter05a,kaliteevski07a}.

Figure~\ref{FigSplitting} shows the dependence of the real parts of the eigen frequencies on the cavity energy $\omega_0$ for different linewidths of the cavity (in meV). The Rabi splitting can be found as a distance between the two branches at the anticrossing point corresponding to $\omega=\omega_0$ and $\omega=\omega_{pl}$. It varies from $\approx 0$ meV ($\gamma=6.5$ meV) to 4.2 meV (for $\gamma=2.2$ meV). Insets illustrate dependences of the splitting on the inverse lifetimes $\gamma$ and $\delta$. Note, that the results presented in Fig.~\ref{FigResponse} and \ref{FigSplitting} are in good agreement with each other.

The main problem from the point of view of experimental observability of the Rabi splitting in such systems is large width of the plasmon resonance \cite{deHeer1993}. However, there exist two main ways to bypass it. The first solution is to use positively charged clusters: it's well known that the optical response of metal clusters has a charge dependence \cite{Ekardt}, and the higher the charge state, the narrower the spectrum. Apparantely, it is important for the experimental observability of the Rabi splitting. Besides, charged clusters are more convenient for the experimental investigations since they are operable with external electric fields. The second way of manipulation of the width is connected with the possibility to use clusters embedded in rare gas matrices. For instance, in case of Na$_9^+$Ar$_{164}$ the plasmon peak seems to be narrower than in case of pure Na-9$^+$ \cite{Reinhard}. By these reasons in current manuscript we consider not only Na-8 neutral cluster, but also positive ions of sodium clusters with closed shells Na-9$^+$, Na-21$^+$ and Na-42$^{++}$. The plasmonic peak can be also become narrower at the lower temperatures. Note, if a cluster is embedded in a dielectric matrix, the refractive index of the matrix can change the interaction between the cluster and the cavity. In fact, the dielectric constant $\varepsilon$ enters the Eq.~(\ref{g}) together with $\varepsilon_0$, thus $g\sim \varepsilon^{-1/2}$ and decreases with $\varepsilon$ being increased. Therefore the Rabi splitting also decreases. 

Figure~\ref{FigClusters} illustrates Rabi splitting dependence on the number of atoms in the cluster. For Na-8 we have small splitting of about 1 meV, notwithstanding for Na-42$^{++}$ cluster the splitting reaches the value of about 7 meV. It is clear that in this manuscript we dealed with the simplest Na clusters. Further theoretical and experimental investigation can be done, and we expect Rabi splitting which ammount to dozens of meV for substantial, big clusters.


\section{Conclusions}

In conclusion, we have analyzed the spectrum of
photoabsorbtion of an individual Na-8, Na-21$^+$ and some others clusters embedded in a single mode
photonic microcavity. We have shown that in the region of the giant
plasmonic resonance the regime of strong coupling between
plasmon and cavity photon can be achieved which manifests itself by
formation of the Rabi doublet and mode anticrossing. A cluster can be considered as a $0D$ object with size comparable to those of the QDs. However, due to the many-body effects it demonstrates Rabi splittings several orders of magnitude larger than other $0D$ quantum objects.


\section{Acknowledgments}
The authors thank V.K. Ivanov for useful
discussions. The work was supported by Rannis "Center of Excellence
in Polaritonics" and FP7 IRSES project "POLAPHEN". R.G.P. thanks the
University of Iceland for hospitality.


\begin{thebibliography}{10}
\bibitem{Thermes} The difficulty of the consideration of the clusters with partially
filled shells comes from the necessity to account for great number
of the transitions between various thermes with different orbital
momenta, which becomes a technically complicated task even for N=8.
The standard way to overcome this difficulty consists in using the
so- called "average therm approximation", which becomes exact for
the clusters with completely filled shells. From the point of view of qualitative results, there is of course no difference between the clusters with fully and partially filled shells.
\end{thebibliography}
\end{document}